\documentclass[twocolumn,showpacs,preprintnumbers,amsmath,amssymb,floatfix]{revtex4}

\usepackage{graphicx}
\usepackage{dcolumn}
\usepackage{bm}
\usepackage{natbib}
\usepackage{epstopdf}

\begin{document}


\title{Coupled Optical Resonance Laser Locking}

\author{Shaun  Burd}
 \email{shaunchristopherburd@gmail.com}
 \altaffiliation[Also with ]{Centre for Quantum Technology, University of KwaZulu Natal, Durban 4041, South Africa}.
\author{Hermann Uys}%
 \email{HUys@csir.co.za}
\affiliation{National Laser Centre, Council for Scientific and Industrial Research, Pretoria 0001, South Africa}%

\date{\today}

\begin{abstract}
We have demonstrated simultaneous laser frequency stabilization of a UV and IR laser, to the same spectroscopic sample, by monitoring only the absorption of the UV laser. 
For trapping and cooling Yb$^{+}$ ions, a frequency stabilized laser is required at 369.95nm to drive the $^{2}S_{1/2}$  $ \rightarrow $ $ ^{2}P_{1/2}$ cooling transition. Since the cycle is not closed, a 935.18nm laser is needed to drive the $^{2}D_{3/2}$ $\rightarrow$ $^{3}D_{[3/2]1/2}$ transition which is followed by rapid decay to the $^{2}S_{1/2}$ state. Our 369nm laser is locked to Yb$^{+}$ ions generated in a hollow cathode discharge lamp using saturated absorption spectroscopy. Without pumping, the metastable $^{2}D_{3/2}$ level is only sparsely populated and direct absorption of 935nm light is difficult to detect.  A resonant 369nm laser is able to significantly populate the $^{2}D_{3/2}$ state due to the coupling between the levels.  Fast re-pumping to the $^{2}S_{1/2}$ state, by 935nm light, can be detected by observing the change in absorption of the 369nm laser using lock-in detection of the photodiode signal. In this way simultaneous locking of two optical frequencies in very different spectral regimes is accomplished.  A rate equation model gives good qualitative agreement with the experimental results. This technique offers improved laser frequency stabilization compared to lasers locked individually to the sample and should be readily applicable to similar ion systems. 
\end{abstract}

\maketitle

\section{\label{sec:Intro}Introduction}

Frequency stabilized lasers are an integral part of many atomic physics experiments. Particularly, in ion trap experiments, the laser frequency must be frequency stabilized to be within a fraction of the natural line width of an atomic transition for effective cooling and state manipulation \cite{Metcalf1999}. Additional stabilized lasers are generally required to prevent shelving of the ion in metastable states. To stabilize each laser, separate spectroscopic references or ultrastable Fabry Perot cavities may be used as long term frequency references. Alternatively one laser can be stabilized and used as a reference to lock additional lasers by using a Fabry Perot to measure the relative frequencies between the lasers \cite{Hensinger2010, Monroe2007}. Traditionally, lasers in the visible or infrared have been locked to molecular or atomic vapour cells using standard spectroscopic techniques. The stabilized laser frequency will generally have to be shifted using external acoustic or electro optical modulation to be sufficiently close to the resonance of interest of the trapped ion. This problem can be eliminated by using ions of the same species as the ions that are to be trapped. Streed et al. demonstrated locking a UV diode lasers to Yb$^{+}$ ions generated in a hollow cathode discharge lamp \cite{Kielpinski2008}.\\
\begin{figure}
	\centering
		\includegraphics[scale=1]{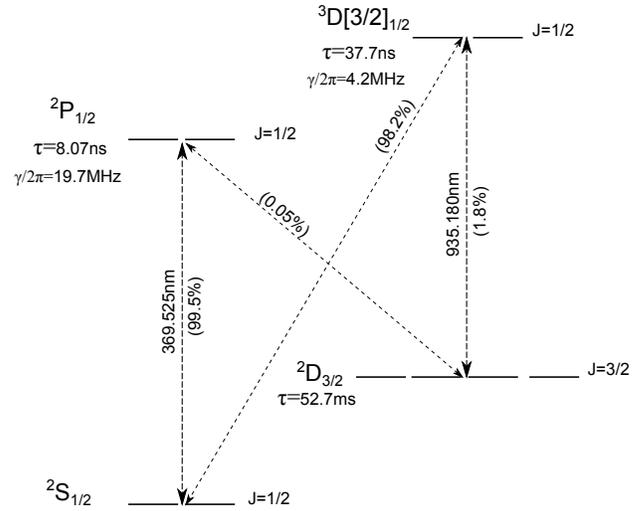}
	\caption{Energy levels of $^{174}$Yb$^{+}$. Spontaneous emission branching ratios are indicated in brackets}
	\label{Fig:Yb174ELevelsS}
\end{figure}
\begin{figure*}
	\centering
		\includegraphics[scale=1]{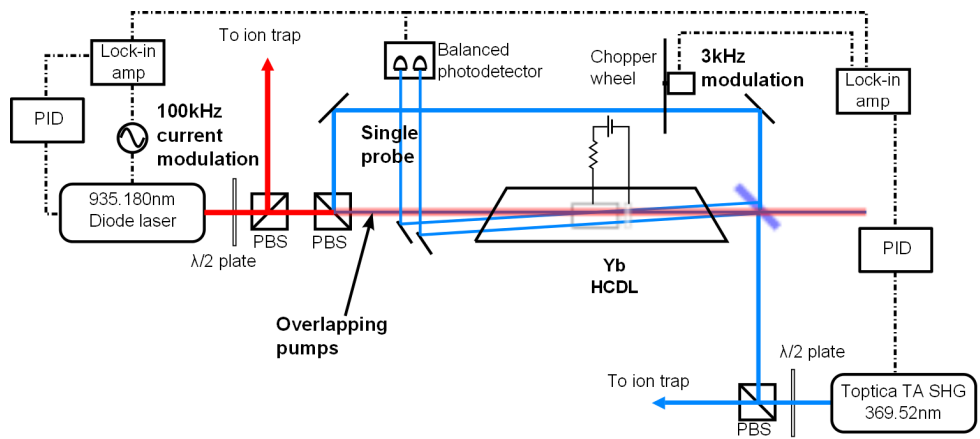}
	\caption{Optical layout for coupled optical resonance laser locking. HCDL - Hollow Cathode Discharge Lamp, PID - Proportional-Integral-Differential controller, PBS - polarization Beam Splitter}
	\label{fig:Layout}
\end{figure*}
Here, we present a method for locking multiple lasers to coupled resonances of ions, in the discharge of a hollow cathode discharge lamp , which we call Coupled Optical REsonance Laser Locking, or CORELL. By allowing lasers with different wavelengths to overlap and interact with the same ion sample, we can obtain separate signals suitable for locking each laser even though the absorption of only one of the lasers is observed. We demonstrated the technique experimentally for $^{174}\text{Yb}^{+}$. Although laser locking to metastable state transitions has previously been demonstrated \cite{Moon2004}, this is the first time, to the best of our knowledge, that multiple lasers have been locked to coupled transitions of ions in single spectroscopic sample by observing the absorption signal from a probe beam of only one of the lasers.

Figure \ref{Fig:Yb174ELevelsS} shows the relevant energy levels of the cooling and main re-pumping transitions of $^{174}\text{Yb}^{+}$ for ion trap experiments. The  $^{2}$S$_{1/2}$ $\rightarrow$ $^{2}$P$_{1/2}$  cooling transition has a center wavelength at 369.5nm. This cooling cycle is not completely closed due to a spontaneous emission branching ratio of 0.005, from the $^{2}$P$_{1/2}$  to the metastable  $^{2}$D$_{3/2}$  state \cite{Monroe2007}. A trapped ion that falls into the metastable $^{2}$D$_{3/2}$ state will no longer interact with the cooling beam. To return the ion to the cooling transition a 935.180nm laser is  used to excite ions from the $^{2}$D$_{3/2}$ state to the $^{3}$D$[3/2]_{1/2}$ state, where they rapidly decay back to the $^{2}$S$_{1/2}$ ground state with 0.982 probability. \\ 

Without optical pumping, few ions in the electric discharge of our discharge lamp will be in the $^{2}D_{3/2}$ state. However, if the $^{2}$S$_{1/2}$ $\rightarrow$    
 $^{2}$P$_{1/2}$ is driven at resonance, there will be a net increase of ions in the $^{2}$D$_{3/2}$ state and a decrease in the population of ions in the ground state. If the 935nm laser is then tuned to the $^{2}$D$_{3/2}$  $\rightarrow$ $^{3}$D$[3/2]_{1/2}$ transition, the metastable state will be depleted and the $^{2}$S$_{1/2}$ state will be repopulated. Thus the 935nm laser will affect the population of the $^{2}$S$_{1/2}$ ground state. The effective lifetime for decay from $^{2}$P$_{1/2}$ into the $^{3}$D$[3/2]_{1/2}$ state is $\tau _{eff}=8.07ns\times0.005=1.6\mu s$. The rate of population transfer to the $^{2}D_{3/2}$ state is therefore 3.3$\times10^5$ greater than the rate of decay of the metastable state directly back to the $^{2}$S$_{1/2}$ state. This large flow of ions into the $^{2}$D$_{3/2}$ state can cause an observable change in the absorption of the 369nm laser. The effect of the separate lasers can be observed with lock-in detection by monitoring the saturated absorption signal of the 369nm laser and modulating each laser at a different frequency.  
    
 \section{\label{sec:Ex}Experimental Setup}

The experimental setup is shown in Figure \ref{fig:Layout}. 369.5nm light is generated by frequency doubling an amplified external cavity 739.05nm diode laser (Toptica Photonics TA-SHG). The 935nm beam is generated using an external cavity diode laser (Toptica Photonics DL 940). The 369nm beam is split into a strong pump beam and 2 weak probe beams by passing it through a Calcium Fluoride plate. The pump and probe beams counter-propagate through an Yb hollow cathode discharge lamp (Hamamatsu L2783-70NE-Yb). The pump beam is made to overlap with one of the probe beams, the other weak beam is used as an unsaturating reference to reduce the Doppler background. After passing through the lamp, the two probe beams are directed to a balanced photodetector (New Focus Model 2107) where they are subtracted to produce a Doppler free saturated absorption signal. To improve the signal to noise ratio, the pump beam is chopped at 3kHz and a lock in amplifier is used to filter and amplify the signal from the balanced photodetector. The 935nm beam propagates collinearlly with the 369nm pump beam. The two beams have orthogonal polarizations and are combined using a beam splitter polarizer cube.

The 935nm spectroscopic signal is also contained in the output signal of the balanced photodetector which only monitors the intensity of the 369nm probe beams. The photodetector signal is used as the input to a second lock-in amplifier. The 935nm beam is modulated at a different frequency from the 369nm laser. Intensity modulation of the 935nm laser beam by chopping can be used to observe the resonance peaks. A derivative signal that is more suitable for laser locking can be generated by using frequency modulation \cite{Demtroder2008}, accomplished by modulating the 935nm diode current.

The wavelengths of the lasers are monitored with a Bristol Model 621 wavelength meter.

\section{\label{sec:Res}Results}

\begin{figure}
	\centering
		\includegraphics[scale=1]{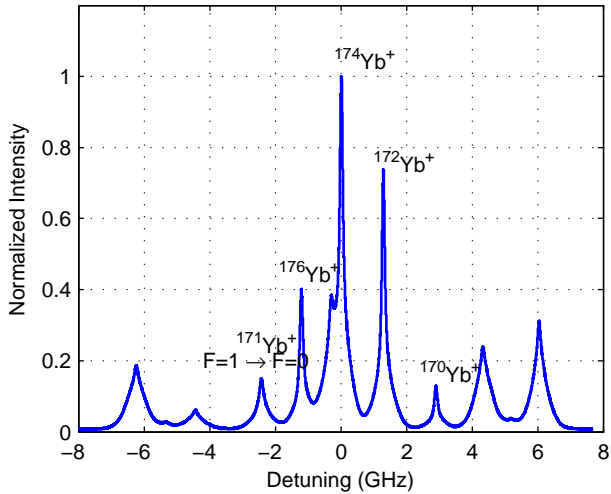}
	\caption{Saturation spectrum of the hyperfine components of the $2$S$_{1/2}\rightarrow 2$P$_{1/2}$ transition of the isotopes of Yb$^{+}$ in a hollow cathode discharge lamp. A lock-in time constant of 3ms was used, and the graph shows the average over 16 oscilliscope traces.}
	\label{fig:YbDFS}
\end{figure}

To observe the hyperfine structure of the $2$S$_{1/2}\leftrightarrow 2$P$_{1/2}$ transition of the Yb$^{+}$ isotopes in the hollow cathode lamp, the frequency of the 369nm laser is scanned from 369.521nm to 396.527nm by adjusting the grating of the 739nm master laser. The spectrum is shown in Figure \ref{fig:YbDFS}. A pump beam power of 10mW and probe power of 0.45mW were used. Precise resonant frequencies for the different ytterbium isotopes for the 369nm transition have been measured by McLoughlin et al.\cite{Hensinger2010} who used trapped ions as a reference. These values correspond closely with the measured wavelengths of resonance peaks in the saturated absorption spectrum and were used to identify the isotopes, as shown in Fig.~3. 


\begin{figure}
	\centering
		\includegraphics[scale=1]{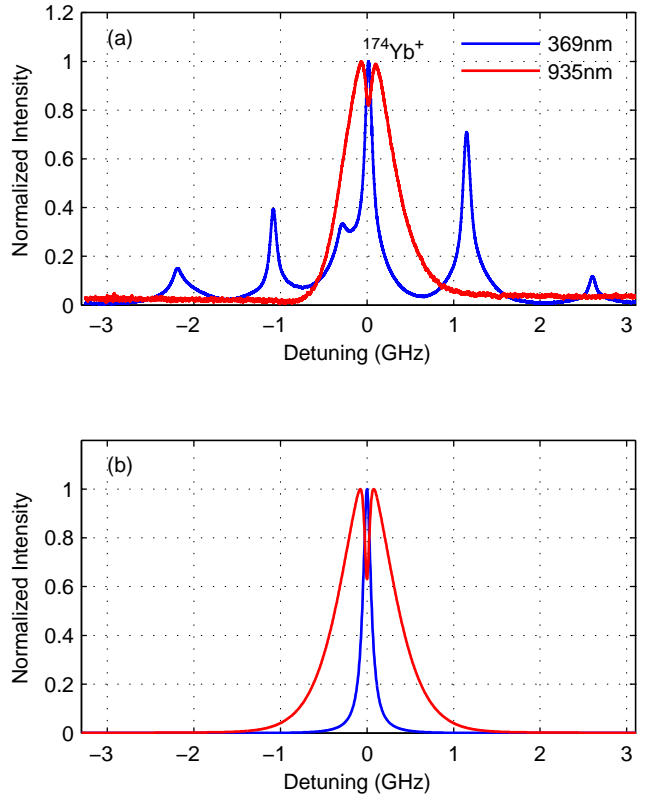}
	\caption{Corell signals obtained by scanning the 369nm laser frequency and holding the 935nm laser on resonance for the repump transition of $^{174}$Yb$^{+}$.  The signal due to the 369nm pump (blue), and due to the 935nm repump (red) are extracted by lock-in detection of the 369nm probe, while modulating the two pumps at different frequencies. Experimental results are shown in (a). The FWHM of the 935nm laser signal is 683MHz. A lock-in time constant of 3ms was used, and the graph shows the average over 16 oscilliscope traces. (b) shows the simulation results of the rate equation model including only the isotope $^{174}$Yb$^{+}$.}
	\label{fig:CorelSlo}
\end{figure}

When the 935nm laser, on resonance with the $^{2}$D$_{3/2}$  $\rightarrow$ $^{3}$D$[3/2]_{1/2}$ transition of a given isotope, interacts with the ions in the lamp, and the 369nm laser is scanned over the $^{2}$S$_{1/2}$ $\rightarrow$  $^{2}$P$_{1/2}$ transition of the same isotope, ions in the $^{2}$D$_{3/2}$ state will be transfered to the $2$S$_{1/2}$ state and there will be increased absorption of 369nm photons. Lock-in detection of the photodetector signal, at the modulation frequency of the 935nm laser, produces an output signal of a broadened peak with a prominent central dip. Figure \ref{fig:CorelSlo} (a) shows the CORELL signals for $^{174}$Yb$^{+}$ observed by lock-in detection at the two different modulation frequencies of the two different lasers. 8.5mW of 935nm light is focused through the lamp and overlaps with the 369nm pump. Observe that the center of the signal due to the 935nm laser occurs exactly at the center to the $2$S$_{1/2}\rightarrow 2$P$_{1/2}$ transition of $^{+}$Yb$^{174}$ and that no CORELL features appear at the modulation frequency of the 935nm laser for the other isotopes. It is important to note that if the frequency of the 935nm laser is slightly detuned from resonance, the magnitude and signal-to-noise ratio of the CORELL signal will reduce, but the center of the feature will still correspond to the peak of the 369nm resonance. The  CORELL signal extracted at the modulation frequency of the 935nm laser shown in Figure \ref{fig:CorelSlo} is a measure of the increase in the absorption of the 369nm laser due to the re-population of the $^{2}$S$_{1/2}$ state by the 935nm laser. This increase in absorption increases the signal to noise ratio of the error signal used to lock the 369nm laser. Figure \ref{fig:CorelSlo} (b) shows the CORELL signals obtained using a rate equation model where only $^{174}$Yb$^{+}$ is considered. The model is discussed in more detail in Appendix A.    


\begin{figure}
\centering
\includegraphics[scale=1]{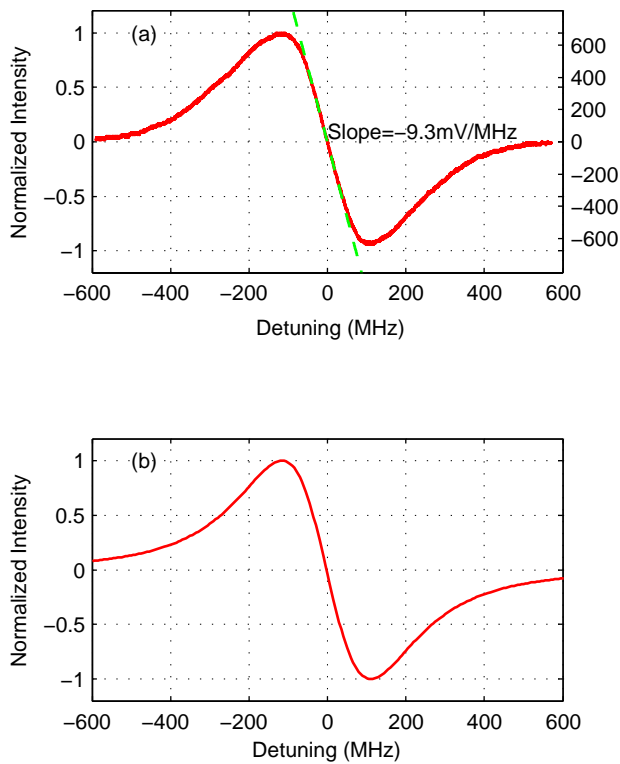}
\caption{Error signal generated by frequency modulation of the 935nm laser while scanning it over the repump transition. The experimentally obtained error signal is shown in (a). The dashed green line indicates the slope at the zero crossing. The simulated error signal obtained using the rate equation model is shown in (b).}
\label{fig:Error}
\end{figure}

\begin{figure}
\centering
\includegraphics[scale=1]{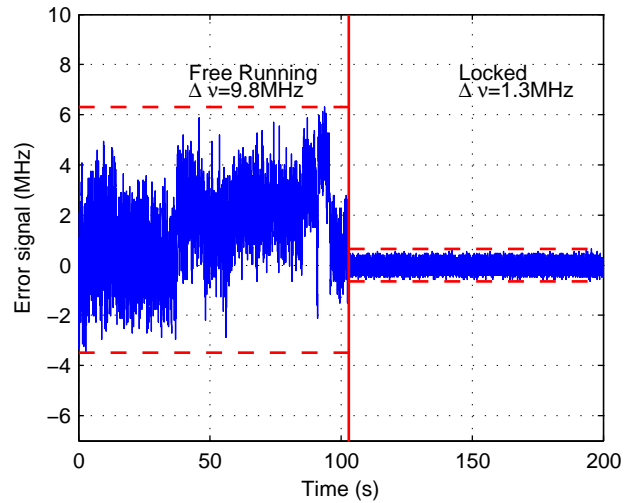}
\caption{Comparison of 935nm laser frequency fluctuations for open and closed loop operation. The solid red line indicates the time at which the laser was locked. The dashed red lines bound the maximum and minimum frequency fluctuations.}
	\label{fig:error_UL_L}
\end{figure}

To generate a signal suitable as a long term reference for frequency stabilization of the 935nm laser, the 369nm laser is locked to the $2$S$_{1/2}\rightarrow 2$P$_{1/2}$ of a Yb$^{+}$ isotope and the 935nm laser is scanned across the $^{2}$D$_{3/2}$  $\leftrightarrow$ $^{3}$D$[3/2]_{1/2}$ of the same isotope. Frequency modulation of the 935nm beam and lock-in detection produce a high signal-to-noise ratio derivative signal. The CORELL signal for $^{174}$Yb$^{+}$ is shown in Figure \ref{fig:Error} (a). The full width at half maximum (FWHM) of the CORELL resonance is measured to be 200MHz using the wavelength meter. The slope of the zero crossing point was measured to be -9.3mV/MHz. Detuning the 369nm laser slightly from resonance reduces the signal to noise ratio, but does not alter the central frequency of the CORELLL signal. With a lock in time constant of 30ms, the maximum observed signal-to-noise ratio is 70. This is significantly improved in comparison to the signal-to-noise ratio that has been obtained using optogalvanic spectroscopy of the repump transition with the 935nm laser alone by Petrasiunas et. al. (2012) who obtained a maximum signal to noise ratio of 58 for a lock in time constant of 300ms using 18mW of 935nm light \cite{Petrasiunas2012}. Figure \ref{fig:Error} (b) shows the error signal predicted by the rate equation model.

The 935nm laser was locked to the zero crossing point of the $^{174}$Yb$^{+}$ CORELL signal shown in Figure \ref{fig:Error}. The frequency stability was estimated using the slope of the zero crossing point of the error signal. Figure \ref{fig:error_UL_L} compares the free running frequency drift of the error signal of the 935nm laser to the frequency fluctuations when the laser was locked using a PID to feedback to the external cavity grating. The CORELL signal provides a long term frequency reference and the closed loop system is able to reject disturbances imposed by changing environmental conditions. The maximum frequency fluctuation observed when the laser was locked was 1.3 MHz which is well within the 4.2MHz linewidth of the repumping transition.

\section{\label{sec:Conclusion}Conclusion}

We have presented a method for simultaneously locking multiple lasers to coupled resonances in a single spectroscopic sample and applied the method for locking separated lasers to the $^{2}$S$_{1/2}\rightarrow ^{2}$P$_{1/2}$  and $^{2}$D$_{3/2}$  $\rightarrow$ $^{3}$D$[3/2]_{1/2}$ transitions of  $^{174}$Yb$^{+}$.  Separate but coupled spectroscopic signals suitable for laser locking can be obtained for all of the lasers by observing only a single probe beam from one of the lasers. The separate signals are distilled from the single probe signal by modulating each laser at a different frequency and using lock-in detection. Due to the coupling between the transitions, the absorption of each laser is increased when the lasers are all on resonance. Thus the technique intrinsically provides improved signal-to-noise ratio for laser locking in comparison to lasers locked to the sample in isolation. This technique should be readily applicable to other ion systems with similarly coupled cooling and repumping transitions. Although we have motivated CORELL for atomic physics applications, it should be very useful in high resolution spectroscopic applications ranging from LIDAR to frequency references in optical communications.

\appendix
\section{\label{sec:Mod} Theoretical Model}

To provide theoretical corroboration of the experimental results a rate equation model was developed. For each of the $K=10$ Zeeman sub-levels shown in Figure \ref{Fig:Yb174ELevelsS}, the time derivative of the fractional population ($n_i$) in level $i$ is
\begin{equation}
\frac{dn_{i}}{dt}=\sum_{j=1}^{K}[P_{ij}(n_{i}-n_{j})+A_{ij}(n_{i}-n_{j})+R_{ij}(n_{i}-n_{j})],
\label{eq:Rate}
\end{equation}
where $A_{ij}$ is the Einstein $A$ coefficient that describes spontaneous emission from level $i$ to level $j$. The elements $R_{ij}$ account for population transfer due to other mechanisms that occur in the hollow cathode discharge. In particular there exists a background population in the metastable $^{2}$D$_{3/2}$ level in the plasma \cite{Petrasiunas2012}.  The $R_{ij}$ elements coupling the $^{2}S_{1/2}$ to $^{2}D_{3/2}$ levels were used as tuning parameters for the model.  $P_{ij}$ describes the coupling between levels $i$ and $j$ due to the laser fields, namely
\begin{equation}
P_{ij}=S_1(i,j)\frac{I_{1}\sigma_{i,j}g(w_{1})}{\hbar \omega_{1}}+S_2(i,j)\frac{I_{2}\sigma_{i,j}g(w_{2})}{\hbar \omega_{2}}.
\end{equation}
Here the subscripts $q=1$ or $2$ denote the 369nm and 935nm lasers respectively.  $\omega_q$ is the angular frequency, and $I_q$  the incident intensity of  laser $q$. $S_{q}(i,j)$  describes the transition strength between levels $i$ and $j$, appropriately weighted by the polarization vector of laser $q$. $\hbar$ is the reduced Planck's constant. The peak value of the absorption cross-section is given by 
\begin{equation}
\sigma_{i,j}=\frac{\lambda_{i,j}^{2}}{2\pi},
\end{equation}
where $\lambda_{i,j}$ is the on-resonance wavelength of the transition $i\rightarrow j$.\\

The line shapes of the lasers depend on the topology of the spectroscopy setup. For the 369nm laser with a pump and a probe beam, the line shape is given by
\begin{eqnarray}
g(\omega_{1})=R_{\text{probe}}\frac{\left( \Gamma_{1}/2\right)^{2}}{\displaystyle \left( \Gamma_{1}/2\right)^{2}+  \left( \omega_{1}-\omega_{1,0}-\textbf{v}_{c} \cdot \textbf{k}_{1} \right)^{2}} \nonumber \\
+ R_{\text{pump}}\frac{ \left( \Gamma_{1}/2\right)^{2}}{\displaystyle \left( \Gamma_{1}/2\right)^{2}+ \left( \omega_{1}-\omega_{1,0}+\textbf{v}_{c} \cdot \textbf{k}_{1} \right)^{2} }\nonumber\\
\end{eqnarray}
$R_{\text{probe}}$ is the fraction of $I_{1}$ in the probe beam, and $R_{\text{pump}}$  the fraction of $I_{1}$ in the pump beam such that $R_{\text{probe}}+R_{\text{pump}}=1$. The numerical values of $R_{probe}=0.04$ and the power of each laser used in the model were the same as the experimental values. $\Gamma_{1}$ is the homogeneous line width of the $^{2}$S$_{1/2}\rightarrow ^{2}$P$_{1/2}$ transition with center frequency $\omega_{1,0}$. The pressure broadening of the line was measured experimentally to be 43$\pm$5MHz. \textbf{v}$_{c}$ is the velocity of ions in velocity class $c$. \textbf{k}$_{1}$ is the wave vector of laser 1. \\

For the 935nm laser co-propagating with the 369nm pump beam
\begin{equation}
 g(\omega_{2})=\frac{\left( \Gamma_{2}/2\right)^{2}}{\displaystyle \left( \Gamma_{2}/2\right)^{2} +  \left( \omega_{2}-\omega_{2,0}+\textbf{v}_{c} \cdot \textbf{k}_{2} \right)^{2}}.
\end{equation}
$\Gamma_{2}$ is the homogeneous line width of the $^{2}$D$_{3/2} \rightarrow$$^{3}D_{[3/2]1/2}$ transition with center frequency $ \omega_{2,0}$. The pressure broadening of this line used as a tuning parameter for the model, since it is not easily measurable in our system. \textbf{k}$_{2}$ is the wave vector of laser 2. \\

The ions are assumed to have a Boltzmann velocity distribution \cite{Demtroder2008} given by
\begin{equation}
f(v_{z})dv_{z}=
\frac{1}{\sqrt{\pi}}e^{-(v_{z}/v_{p})^{2}}dv_{z},
\end{equation}
where $f(v_z)$ is the fraction of ions with velocity between $v_{z}$ and $v_{z}+dv_{z}$. $v_{p}$ is the average velocity. To calculate the absorption spectrum, the ions are divided into discrete velocity classes with fractional weighting $f_{l}$ described by the Boltzmann distribution. At the given laser frequencies $\omega_1$ and $\omega_2$, steady state level populations ($n_i(l,\omega_{1},\omega_{2})$) are calculated from the rate equations (Equation \ref{eq:Rate}) for each velocity class. The total population in each level for the entire sample is then given by

\begin{equation}
N_i(\omega_{1},\omega_{2})=\sum_{l}f_{l}n_{i}(l,\omega_{1},\omega_{2})
\end{equation}
The absorption coefficient for the 369nm probe beam can then be calculated by summing over all velocity classes:

\begin{eqnarray}
\alpha(\omega_{1},\omega_{2})=\sum^{K}_{i}\sum^{K}_{j}S_{1}(i,j) \times  \nonumber \\
\sum_{l}g_{1}(\omega_{1},l)[n_{i}(l,\omega_{1},\omega_{2})-n_{j}(l,\omega_{1},\omega_{2})]N\nonumber \\
\end{eqnarray}

where and $N$ is the total number density of $^{174}$Yb$^{+}$ ions in the hollow cathode discharge lamp.  

\begin{equation}
g_{1}(\omega_{1},l)=\frac{\left( \Gamma_{1}/2\right)^{2}}{\displaystyle \left( \Gamma_{1}/2\right)^{2}+  \left( \omega_{1}-\omega_{1,0}-v_{l} k_{1} \right)^{2}}
\end{equation}

The photodiode signal is proportional to power absorbed by the laser which is exponentially related to the absorption coefficient. Thus we can obtain a normalized model of the photodiode signal $D$ from: 
\begin{equation}
D(\omega_{1},\omega_{2})\propto e^{-\alpha(\omega_{1},\omega_{2})L},
\end{equation}
where $L$ is the length of the sample 

\bibliography{References}

\end{document}